\documentclass[prb,aps, twocolumn, amsmath,amssymb,superscriptaddress]{revtex4}
\usepackage{float}
\usepackage{amsmath}
\usepackage{amssymb}
\usepackage{amsfonts}
\usepackage{euscript}
\usepackage{enumerate}
\usepackage{hhline}
\usepackage{pslatex}
\usepackage{tabularx}
\usepackage{threeparttable}
\usepackage[usenames,dvipsnames]{xcolor}
\usepackage[colorlinks,linkcolor=red,anchorcolor=blue,citecolor=blue]{hyperref}

\usepackage{graphicx}% Include figure files
\usepackage{dcolumn}% Align table columns on decimal point
\usepackage{bm}% bold math
\usepackage[sort&compress]{natbib}
\usepackage{balance}

\makeatletter

\begin{document}

\title
{Low-threshold nanolasers based on miniaturized bound states in the continuum}

\author{Yuhao Ren}
\thanks{These authors contributed equally}
\affiliation{State Key Laboratory of Optoelectronic Materials and Technologies, School of Physics, Sun Yat-sen University, Guangzhou 510275, China}

\author{Peishen Li}
\thanks{These authors contributed equally}
\affiliation{State Key Laboratory of Advanced Optical Communication Systems and Networks, School of Electronics \& Frontiers Science Center for Nano-optoelectronics, Peking University, 100871, Beijing, China}

\author{Zhuojun Liu}
\affiliation{State Key Laboratory for Mesoscopic Physics and Frontiers Science Center for Nano-optoelectronics, School of Physics, Peking University, 100871, Beijing, China}

\author{Zihao Chen}
\affiliation{State Key Laboratory of Advanced Optical Communication Systems and Networks, School of Electronics \& Frontiers Science Center for Nano-optoelectronics, Peking University, 100871, Beijing, China}

\author{You-Ling Chen}
%\thanks{ylchen@semi.ac.cn}
\affiliation{State Key Laboratory on Integrated Optoelectronics, Institute of Semiconductors, Chinese Academy of Sciences, Beijing 100083, China}

\author{Chao Peng}
%\thanks{pengchao@pku.edu.cn}
\affiliation{State Key Laboratory of Advanced Optical Communication Systems and Networks, School of Electronics \& Frontiers Science Center for Nano-optoelectronics, Peking University, 100871, Beijing, China}
\affiliation{Peng Cheng Laboratory, Shenzhen 518055, China}

\author{Jin Liu}
%\thanks{liujin23@mail.sysu.edu.cn}
\affiliation{State Key Laboratory of Optoelectronic Materials and Technologies, School of Physics, Sun Yat-sen University, Guangzhou 510275, China}

\date{\today}% It is always \today, today,

\begin{abstract}
The pursuit of compact lasers with low-thresholds has imposed strict requirements on tight light confinements with minimized radiation losses. Bound states in the continuum (BICs) have been recently demonstrated as an effective mechanism to trap light along the out-of-plane direction, paving the way to low-threshold lasers. To date, most reported BIC lasers are still bulky due to the absence of in-plane light confinement. In this work, we combine BICs and photonic band gaps to realize three-dimensional (3D) light confinements, as referred to miniaturized (mini-) BICs. Together with 3D carrier confinements provided by quantum dots (QDs) as optical gain materials, we have realized highly-compact active BIC resonators with a record-high quality ($Q$) factor up to 32500, which enables single-mode continuous wave (CW) lasing with the lowest threshold of 80 W/cm$^{2}$ among the reported BIC lasers. In addidtion, our photon statistics measurements under both CW and pulsed excitations confirm the occurence of the phase transition from spontaneous emission to stimulated emission, further suggesting that conventional criteria of input-output and linewidth are not sufficient for claiming nanoscale lasing. Our work reveal a via path towards compact BIC lasers with ultra-low power consumption and potentially boost the applications in cavity quantum electrodynamics (QEDs), nonlinear optics and integrated photonics.

\noindent{\textbf{keywords: }Nanolaser, Bound states in the continuum, Photon statistics.}%Use showkeys class option if keyword

\end{abstract}

\maketitle

\section{Introduction}
Nanoscale coherent light generations via stimulated emissions have been the scientific frontier of nanophononics, topological photonics\cite{lu2014topological,ozawa2019topological,ota2020active}, non-Hermitian physics\cite{hodaei2014parity,peng2014loss,ozdemir2019parity} and optics in random media\cite{wiersma2008physics,sapienza2019determining}. From the viewpoint of technology, the scalable creations of miniaturized lasers with low-power consumption enable a variety of important applications across optical interconnects\cite{tucker2010green,tatum2015vcsel}, bio-sensing\cite{wang2017lasing} and far-field beam synthesis \cite{bahari2017nonreciprocal,zhang2020ultrafast} etc. To achieve lasing at the extreme sub-wavelength scale, plasmonic cavities are usually employed however they unavoidably suffer from high Ohmic losses associated with metals\cite{wang2020loss,khurgin2014comparative}. While at the wavelength scale,  dielectric nanolasers have been realized with the assistance of high-quality ($Q$) cavities utilizing total internal reflection or photonic bandgaps (PBGs)\cite{kim2012graphene,painter1999two,imada1999coherent}, such as micro-disks or photonic crystal (PhC) defect cavities. However, due to the limited lasing volume, their emission power is still not quite sufficient in driving the applications, for instance, on-chip optical communications. Recently, several new designs such as random laser\cite{lee2019taming,trivedi2022self}, topological laser\cite{smirnova2020room,shao2020high} and moire lattice laser\cite{mao2021magic} had been proposed to achieve lasing behavior at ten-wavelength scale, to best compromise the footprint and power.

%Beyond wavelength scale, topologically protected edge states or multiple scattering in random media could be explored to build robust and reconfigurable laser devices against disorders\cite{shao2020high,smirnova2020room,trivedi2022self,lee2019taming}.

Trapping light is no doubt the first step towards nanoscale lasers. As an emerging mechanism, BICs have been demonstrated as a very powerful tool to suppress out-of-plane radiations and consequently boost the Q-factors of planar optical resonators\cite{hsu2013observation,hsu2016bound}. In addition, the vectorial nature of BICs enables emissions of structured light from chip-scale devices, leading to ultra-fast switchable nanolasers \cite{huang2020ultrafast} and multiplexed nanolasers carrying orbital angular momenta\cite{bahari2021photonic}.
In principle, ideal BICs with infinite $Q$s only exist in periodic and symmetric structures. Therefore, early demonstrations of BICs employed relatively large sample sizes ranging from a-few-tens to hundreds of periodic unit-cell in order to maintain the high-$Q$ feature\cite{kodigala2017lasing,ge2019laterally}. One of the successful efforts in promoting the $Q$s is to topologically merge a set of BICs into the so-called super-BIC regime \cite{jin2019topologically}, which dramatically minimizes the radiation loss and therefore significantly reduces the thresholds of BIC lasers \cite{hwang2021ultralow}.

To make the nanolaser suitable for the practical applications, we look for the BICs in miniaturized sizes, namely ten-wavelength scale. For active devices, the benefits of shrinking the mode sizes lies on two folds: first, in the spontaneous emission regime, the small mode volumes can greatly enhance the strength of light-matter interactions at single-photon level, enabling the explorations of cavity QED effects in both weak- and strong-coupling regimes\cite{gerard2003solid}; second, in the stimulated emission regime, miniaturized mode volumes produce strong light trapping which significantly reduces the lasing thresholds\cite{ma2019applications}.
However, the symmetry breaking or truncating the infinite size, transiting ideal BICs to quasi-ones\cite{koshelev2018asymmetric,liu2019high}, unavoidably lowers down the $Q$-factors accordingly and makes them less favorable for low-threshold lasing. Thus it is highly-desirable yet an on-going challenge to achieve mini-BICs with high-$Q$ factors.

%The introduction of symmetry breakings or truncation of the infinite size can trigger the transition from BICs to quasi-BICs with finite Q-factors for practical applications\cite{koshelev2018asymmetric,liu2019high}. In general, the less deviations from the ideal BICs, the higher Q-factors can be achieved for the quasi-BICs. Therefore. early demonstrations of BIC-based devices usually employ relatively-large sample sizes from a few tens to hundreds of periodic arrangements of the unit cell in order to maintain high-Q factors. It is highly-desirable yet an on-going challenge to achieve mini-BICs with high-Q factors. The benefits of shrinking the sizes of BICs for active devices lies on two folds. In the spontaneous emission regime, the small mode volume can greatly enhance the strength of light matter interactions at single-photon level, enabling the explorations of cavity QED effects in both weak- and strong coupling regimes\cite{gerard2003solid}. In the stimulated emission regime, mini-BICs produce more compact lasing devices with significantly reduced thresholds\cite{ma2019applications}. One of the successful efforts along this direction is to merge different BICs into the so-called super-BIC regime\cite{jin2019topologically},which dramatically minimizes the radiation loss in finite structures and therefore significantly reduces the thresholds of compact BIC lasers\cite{hwang2021ultralow}.However,appreciable in-plane scattering losses on the boundary of super-BICs prevent further optimizations of the Q-factors, limiting the lasing performances.

\begin{figure}[ht!]
	\begin{center}
		\includegraphics[width=1\linewidth]{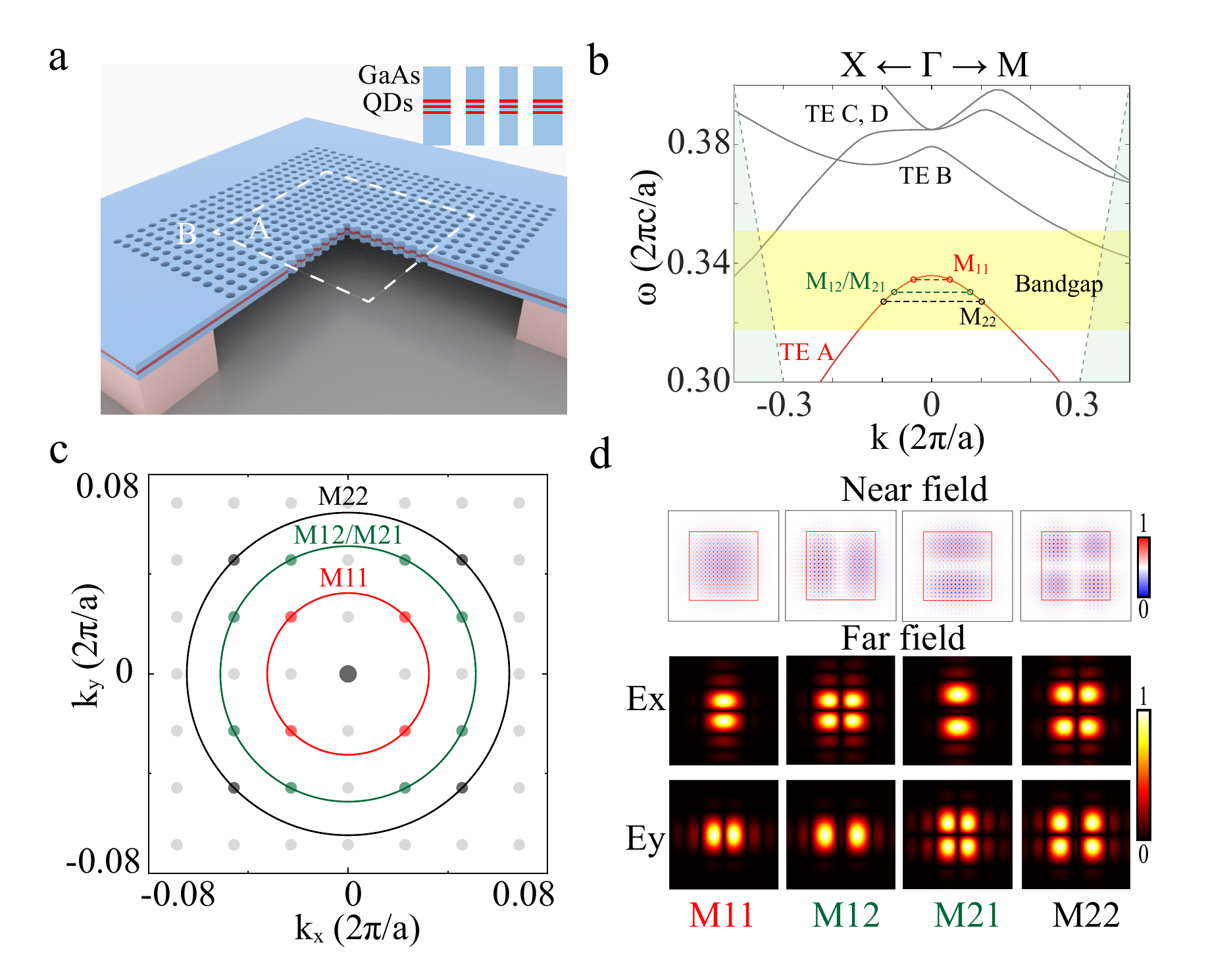}
		\caption{\textbf{Design of mini-BIC lasers.}  a, Schematic of mini-BIC laser device consisting of a suspended GaAs thin membrane with periodically etched air holes. The PhCs in regime A is surrounded by another heterogeneous PCs in region B. Three layers of QDs are embedded in the center of the membrane as optical gain materials. Inset: cross-section of the etched membrane with three layers of QDs. b, The band diagrams of PhCs in regime A (with infinite size). The continuous band (TE A, represented by the red line) associated with ideal PhCs in regime A was quantized into discrete modes above the light line and located in the bandgap of PhCs in region B (represented by the yellow area). c, The momentum distribution of each mode, modes are labeled as M$_{pq}$, according to their momentum peak positions in the first quadrant. d, The near-field ($6~\rm{\mu m} \times 6~\rm{\mu m}$) and far-field patterns ($30^{\circ} \times 30^{\circ}$) of four modes M$_{11}$ through M$_{22}$.}
		\label{fig:Fig1}
	\end{center}
\end{figure}

In this work, we simultaneously employ the BICs to suppress the out-of-plane radiation and utilize the PBGs to minimize the in-plane optical disspations for achieving high-$Q$ optical resonators with small footprints. We fabricated active GaAs membranes supporting mini-BICs with $Q$s as high as $\sim$32500 and exploited high-density InAs QDs as optical gain materials. Laser oscillations at telecom O-band under both CW and pulsed optical pumping were observed with a threshold down to 80 W/cm$^{2}$, which is nearly two order of magnitude lower than the previous reported BIC lasers\cite{hwang2021ultralow}. We systematically compare the lasing and non-lasing behaviors by using time-resolved and photon statistics measurements, further revealing the phase transition from spontaneous to stimulated emission in our devices. This work serves as a crucial step towards understanding and realizing high-performance BIC lasers with miniaturized footprints and may further boost the applications of BICs in cavity QED and integrated nonlinear photonics which demand for both high-$Q$ and low mode volume cavities.

\begin{figure*}[htb!]
	\begin{center}
		\includegraphics[width=1.0\linewidth]{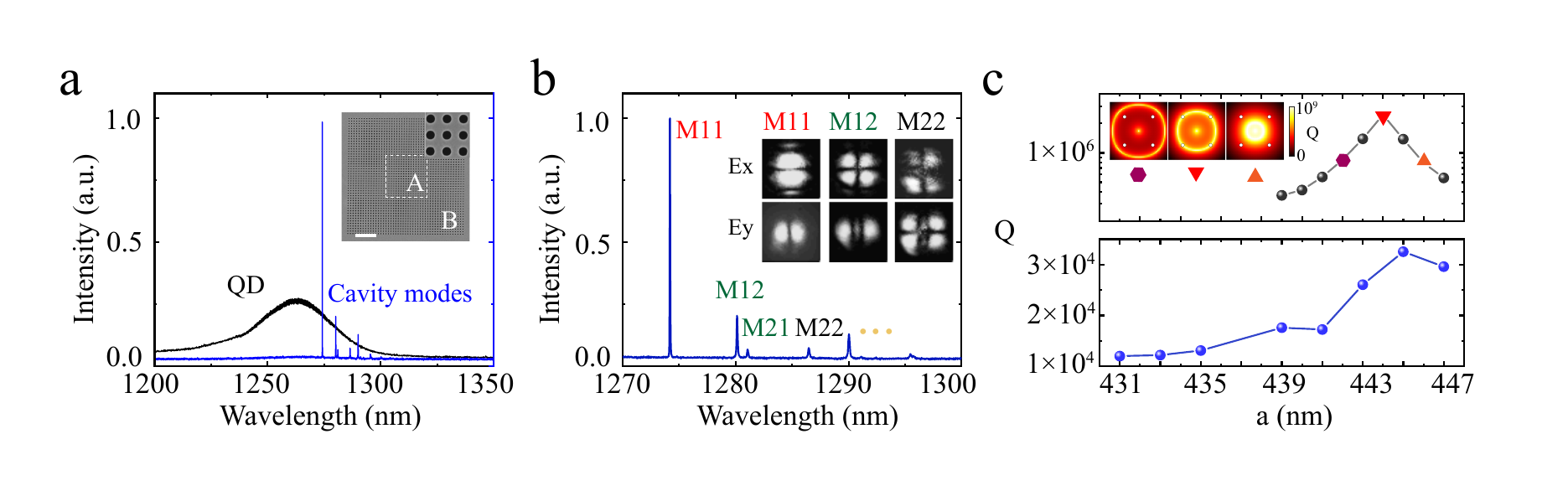}
		\caption{\textbf{Optical characterizations of the active mini-BIC cavity.}  a, $\mu$-PL spectra of QD ensemble (black) and the cavity modes (blue) of the mini-BIC. Inset shows top-view SEM image of the fabricated mini-BIC device. b, Zoomed-in $\mu$-PL spectrum of the cavity modes associated with mini-BICs. The inset shows experimental far-field patterns (x/y-polarized) of modes M$_{11}$ through M$_{22}$. c, Simulated (upper) and measured (lower) $Q$-factors of the M$_{11}$ mode as a function of the lattice constant $a$. Inset: high-Q rings corresponding to the constellation of multiple BICs in momentum space for three different lattice constants.}
		\label{fig:Fig2}
	\end{center}
\end{figure*}

\section{Design}
Our laser cavity is based on PhCs consisting of a suspended GaAs membrane with periodically etched air holes in a square lattice array, as schematically shown in Fig.~1(a). Three layers of high-density ($10^{10}/\rm{cm}^{2}$) InAs QDs are embedded in the center of GaAs membranes as optical gain materials at telecom O-band. The details of the QD epitaxial wafer are presented in Fig.~S1 of the supplementary information (SI). To tightly localize the light in three dimensions, we explored a recent proposal of mini-BICs in which the out-of-plane radiation is suppressed by BICs while the in-plane light confinement is achieved by PBGs \cite{chen2022observation}. The BICs associated with the PhCs in region A (lattice constant $a$) is surrounded by heterogeneous PhCs in region B (lattice constant $b$) with a gap size of $g$ in between. The band diagram of the designed structure is shown in Fig.~1(b), in which the TE-A mode of PhCs  in region A sits in the band gap of PhCs in region B. Therefore, region B serves as highly reflective mirrors to suppress the in-plane light leakage from the mini-BICs residing in region A, and thus significantly improves the $Q$-factor and reduces mode volume $V$. In addition, the finite size of the cavity region A quantizes the continuous TE-A band into discrete modes with a mode spacing of $\delta k$ = $\pi/L$ , where $L$ is the cavity length of region A. Each mode can thus be labeled by a pair of integers $(p, q)$, indicating that its momentum is mostly localized near $p \pi/L$; $q \pi/L$ in the first quadrant of the momentum space. In Fig.~1(c), we plot the highly momentum-localized modes of M$_{11}$, M$_{12}$/M$_{21}$ and M$_{22}$ in the first quadrant of momentum space, in which M$_{12}$ and M$_{21}$ are degenerated in frequency due to the 90-degree rotation symmetry of the structure ($C_4$). The near-field mode distributions reveal that the mini-BICs are highly spatially localized in the cavity region enclosed by the PBG mirrors, while such momentum space localizations result in highly-directional emissions towards specific angles, as presented in Fig.~1(d).

\section{Device fabrication and characterizations}
Experimentally, the mini-BIC patterns were defined on the electron beam (E-beam) resist by using a 100 kV E-beam writer and then transferred to the GaAs layer via a chloride-based dry etch process. The GaAs membranes were released by selectively wet etching 1500-nm-thick AlGaAs sacrificial layer underneath. The full fabrication flow of the devices is presented in Fig.~S2 of SI. Sharp resonances corresponding to the cavity modes were identified by measuring micro-photoluminescence ($\mu$-PL) from the cavity area at room temperature. The details of the home-made $\mu$-PL setup is presented in Fig.~S3 of SI. For comparison, the $\mu$-PL of QD ensemble, indicating the optical gain spectrum, was obtained from the area without any cavity structures. As presented in Fig.~2(a), by choosing a proper lattice constant $a$ = 445~nm with 17 perodicities, $b$ = 463~nm with 15 perdicities and $g$ = 455~nm, cavity modes are spectrally aligned with the gain spectrum to facilitate the lasing oscillations. The sharp resonances measured from the cavity area were further zoomed in Fig.~2(b), where the optical characteristics such as resonant wavelength and mode spacing agreed very well with the simulations in Fig.~1. In additions, we performed the far-field characterizations of each mode, further revealing the vectorial nature and momentum localization of the mini-BIC modes, as predicted in the Fig.~1(d). The dependence of the $Q$-factor (measured at the transparent excitation power) on the constant a of the M$_{11}$ mode is presented in Fig.~2(c). As shown in the simulations, the $Q$s of mini-BIC can be engineered  by tailoring the topological charges of the BICs via tuning the latticed constant of etched air holes.

We measured the $\mu$-PL spectra of one cavity under CW excitation with different excitation powers, as prseented in Fig.~3(a). Sharp cavity modes on a broad emission background can be identified under low excitation powers. By increasing the excitation power, M$_{11}$ mode dominated the emission spectrum and its line-width reduced significantly. The intensity and line-width of the cavity modes as a function of the excitation power were plotted in Fig.~3(b). A sharp increase in the input-output (IO) curve together with reduction of the cavity line-width, commonly believed as signatures of lasing, was observed, suggesting the occurrence of lasing oscillation.

To reliably identify the lasing oscillation, we characterized photon statistics of the emitted photons by measuring the second-order coherence\cite{kreinberg2017emission,jagsch2018quantum}, which can rigorously quatify
the quantum nature of the phase transition from spontaneous emission to stimulated emission. In Fig.~3(c), at very low excitation powers, the emission was in a thermal state but only exhibited a slight bunching behavior in g$^2$(0) due to the relatively short coherence time of the emitted photons. With further increase of the excitation powers, the coherence time got significantly prolonged and therefore appreciable photon bunching effects were observed. As long as the excitation power crossed the lasing threshold of 41~$\rm{\mu}$W with a beam diameter of 8.11 $\mu \rm{m}$ (see Fig.~S4), corresponding to a power density of 80 W/cm$^2$, the thermal state evolved towards a coherent state and the g$^2$(0) gradually lowered down to $1$ when far-above threshold. The photon statistics evidence together with the I-O curve and line-width reduction unequivocally demonstrated the realization of lasing oscillation in our mini-BIC devices.

On the other hand, there have been increasing debates on using the kink feature upon I-O curves as the single criterion of nanoscale lasing, especially when the
device experiences cavity QED effects\cite{samuel2009recognize,reeves20182d,ning2013laser}. To that end, we characterized a less optimal mini-BIC device similarly and presented the experimental data in Fig.~3(e-h). This non-lasing device also exhibited a ``threshold" feature upon the I-O curve as well as a line-width narrowing behavior, which are very similar to the lasing device shown in Fig.~3(a-b). However, no signatures of phase transition from thermal emission to a coherent state were observed from the photon statistics measurements. As shown in Fig.~3(f), the g$^2$(0) raised monotonously with the increase of the excitation power, which indicates that the device was operating as a nano-light-emitting-diode (nano-LED) instead a nanolaser. 
\begin{figure*}[htb!]
	\begin{center}
		\includegraphics[width=0.9\linewidth]{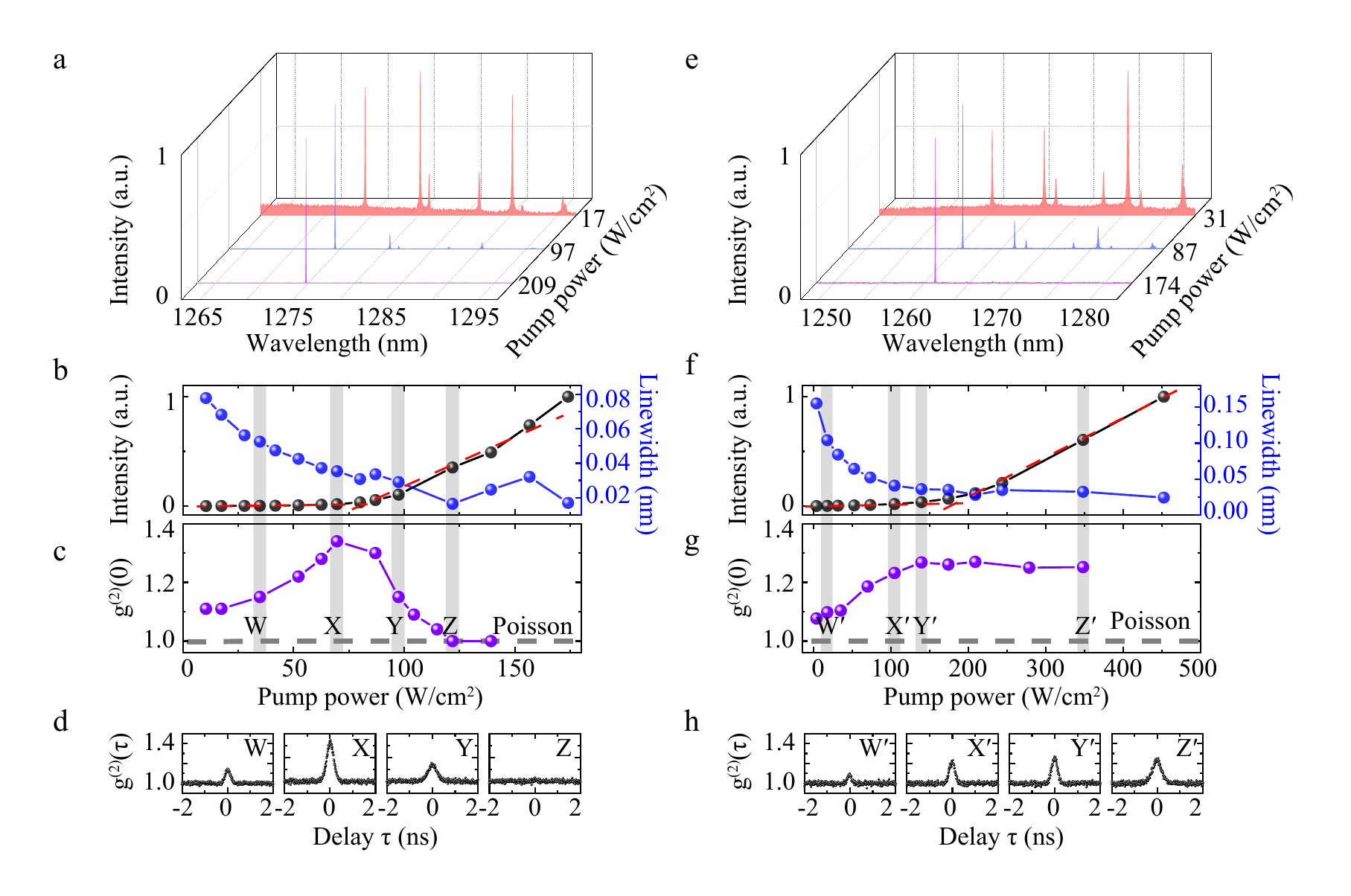}
		\caption{\textbf{Characterizations of lasing and non-lasing behavior under CW excitations.}  a,e, Evolution of the normalized emission spectra of the lasing device (a) and non-lasing device (e) with the increased excitation power. b,f, Integrated output intensity and line-width of the cavity mode M$_{11}$ as a function of the excitation power, showing "threshold-like" behaviors at 80 W/cm$^2$ (b) and 184 W/cm$^2$. The red dashlines shows different slopes the output power when inceasing the excitation power. The shaded areas represnet the excitatiton powers at which the g$^2$ trace are presented. (f), respectively. c,g, Second-order correlation functions at zero delay time g$^2$(0) as a function of the excitation power. The lasing device exhibited a clear phase transition from spontaneous emission (a thermal state with g$^2$(0)$>$1) to stimulated emission (a coherent state with g$^2$(0)=1) while the non-lasing emissions remained in a thermal state for all the excitation powers. d,h, Auto-correlation traces taken at the pump power densities marked in (b,f)}
		\label{fig:Fig3}
	\end{center}
\end{figure*}

Indeed, it is not surprising that the non-lasing device can also exhibit nonlinear output intensities upon the excitation power. Such a fact has been observed in a coupled single-QD cavity system without involving lasing oscillations, as a consequence of the non-resonant couplings between different excitonic states of QDs and the cavity mode\cite{winger2009explanation}. At low excitation powers, the intensity of the cavity mode followed the linearly increased emission of exciton state formed in the singe QD. Under high excitation powers, the cavity mode showed a super-linear power-dependence owing to the non-resonant couplings to the bi-exciton states whose intensities grow nearly as twice as the exciton states. Such a mechanism could also be responsible for the nonlinear output intensities of the many QDs-cavity coupled system, as observed in Fig.~3(d). The line-width narrowing shown in Fig.~3(e) is easier to be understood since the active cavity experienced more absorption from the gain material under low excitation powers. With the increase of excitation power, the excited states of QDs get populated, resulting in the reduction of the cavity line-width.

\begin{figure*}[htb!]
	\begin{center}
		\includegraphics[width=0.9\linewidth]{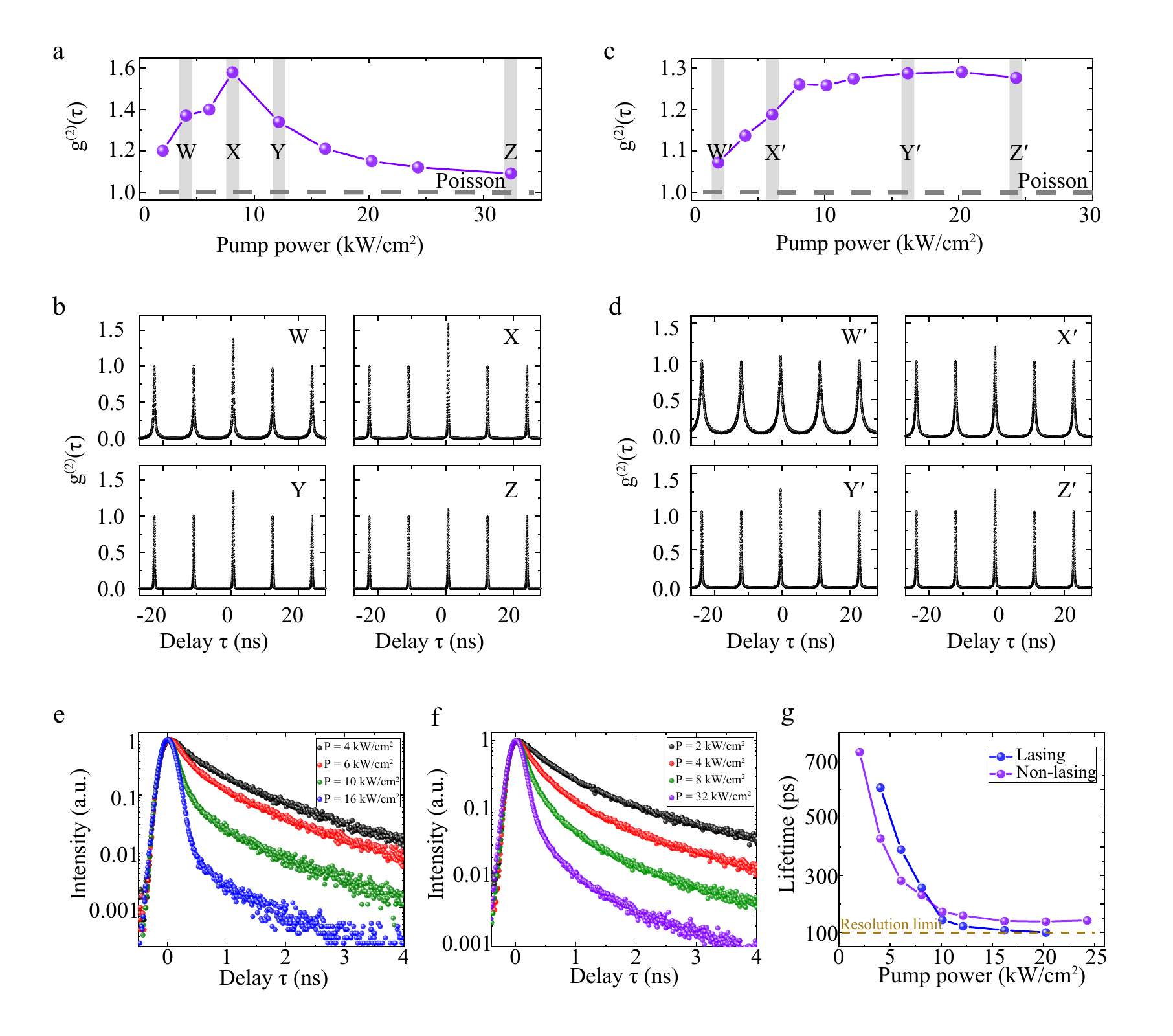}
		\caption{\textbf{Characterizations of lasing and non-lasing devices under pulsed excitation.}  a, Second-order correlation function at zero delay time  g$^2$(0) as a function of the excitation power. The lasing device exhibited a clear phase transition from spontaneous emission (a thermal state with  g$^2$(0)$>$1) to stimulated emission (a coherent state with g$^2$(0)=1).  Auto-correlation traces taken at the pump power densities marked in (a). c, Second-order correlation function at zero delay time  g$^2$(0) as a function of the excitation power. The non-lasing device remained in a thermal state (g$^2$(0)$>$1) through all the excitation powers.  The shaded areas W X Y Z represnet the excitatiton powers at which the g$^2$ trace are presented. (d), Auto-correlation traces taken at the pump power densities marked in (c). e,f The photon lifetime traces of lasing and non-lasing devices under different excitation powers. g, The lifetimes of lasing and non-lasing devices as a function of the excitation power. The orange dash line indicates the time resolution of our measurement system.}
		\label{fig:Fig4}
	\end{center}
\end{figure*}

To investigate the possibility of turning the non-lasing behavior to a lasing oscillation with higher-peak powers and less thermal effects, we further performed pulsed excitation on the same devices. The CW lasing device also exhibited clear a nonlinear IO curve and line-width reductions, as presented in Fig.~S5(a,b) of SI. The pulsed g$^2$(0) clear exhibited the phase transition from a thermal state to a coherent states when increasing the excitation power, as shown in Fig.~4(a,b). On the contrary, the nano-LED device showed ``laser-like" behaviors in terms of nolinear incerease in the I-O curve and line-width reductions, but its photon statistics remained in a thermal state across all the excitation power, as shown in Fig.~4(c,d). One of the advantages of pulsed excitation scheme is that we are able to directly measure the photon lifetime\cite{atlasov2009photonic} which can serve as an additional evidence of stimulated emissions. The decay curves under different excitation powers for the laser and LED devices are presented in Fig.~4(e,f), respectively. Their emissions lifetimes were shortened during the increasing of pulse power, as quantitatively presented in Fig.~4(g). For the lasing device, the lifetime reduced more rapidly than those of the nano-LED till the measurement was beyond the time resolution of our photon-detectors. Such fast decay rates above threshold are strong indications of stimulated emissions. The lifetime of photons emitted by the nano-LED device also decreased, which is expected due to the different carrier relaxation dynamics under high excitation powers\cite{schwab2006radiative}. However, the photon lifetime of nano-LED reduced rather slowly as compared to the lasing device and became saturated (not reaching the time resolution limit of our measurement) at high excitation powers, which is more preferably to ascribe to spontaneous emission.

\section{Conclusion}
We have demonstrated III-V active mini-BIC resonators with a quality factor as high as $\sim$32500, which enables the realizations of lasing oscillations with a record-low threshold of 80 W/cm$^2$ that is nearly two orders of magnitude lower than the state-of-the-art BIC lasers (see table I in SI for comparison). Both CW and pulsed lasing were unequivocally realized by systematically measuring the device characteristics including $\mu$-PL spectra, I-O curves, emission line-width, photon statistics and photon lifetimes. Our investigations suggest that any claim of nanolasing with cavity QED effects should be very careful and further comparisons between more advanced theory and experiments are indispensable. Moving forwards, the vectorial nature of mini-BICs could be exploited to build chip-scale lasers capable of emitting coherent structured light for high-capacity optical communication\cite{willner2021orbital} and high-dimension quantum information processing\cite{erhard2018twisted}. It is also highly desirable to implement the electrical injections in active BIC devcies, as successfully demonstrated for PhC defect and surface emitting lasers in similar structures\cite{ellis2011ultralow,morita2021photonic,itoh2022high}, From the perspective of applications, the realization of mini-BIC laser may immediately boost the development of on-chip cavity QED\cite{vuckovic2014quantum} as well as the integrated nonlinear photonics\cite{hendrickson2014integrated} in which high-Q factors and small mode volumes are highly beneficial.

\noindent \textbf{Acknowledgements}
This research was supported by National Key Research and Development Program of China (2018YFA0306103); National Natural Science Foundation of China (11874437, 62035017, 61922004, 62135001), Major Key Project of PCL (PCL2021A14) and the national super-computer center in Guangzhou.

%\appendix

%\section{Device fabrication }

\noindent \textbf{Device fabrication:} 
The samples are fabricated using a 500 nm-thick GaAs/1.5~$\mu$m-thick Al$_{0.8}$Ga$_{0.2}$As/GaAs substrate wafer. Three layers InAs/GaAs QDs are embedded in the middle of the first GaAs layer, whose central emission wavelength is ~1.3~$\mu$m. The AlGaAs layer act as sacrificial and etch stop layer. First, a layer of Si$_3$N$_4$ with a thickness of 200 nm was deposited on the wafer by Inductively coupled plasma chemical vapor deposition (ICP-CVD) as a hard mask for PhC dry etching. A 400~nm ARP6200 electron beam resist was spin coated on the surface of the hard mask. Subsequently, the electron beam lithography was used to define the PhC pattern in ARP6200. The PhC pattern was transferred from resist into the hard mask layer using reactive ion etching (RIE). Afterwards, the electron beam resist was removed by Inductively coupled plasma etching (ICP) with O$_2$ plasma. Then ICP etching was performed to obtain the air holes through the active layer. The residual Si$_3$N$_4$ hard mask was removed by RIE dry etching. Finally, the sacrificial layer was wet etching by immersing the sample in 10$\%$ hydrofluoric acid solution for 1min30s to form an air region slab underneath the PhC membrane.

\noindent \textbf{Optical measurement:} 
The fabricated mini-BIC devices were characterized with a customized micro-photoluminescence ($\mu$-PL) measurement system in a surface-normal pump configuration. 780 nm CW laser diode and pulse laser diode (10 ps, 86~MHz period) were used to optically pumped the sample via a 50x objective lens with a numerical aperture of 0.65. The emission spectra were collected by the same objective and analyzed by a spectrometer (iHR550) with a resolution of 0.025 nm.

%\bibliographystyle{elsarticle-num-names}

%\bibliography{Ref1}

\balance

\newpage
\onecolumngrid \bigskip

\begin{center} {{\bf \large SUPPLEMENTARY INFORMATION}}\end{center}

\setcounter{figure}{0}
\setcounter{section}{0}
\makeatletter
\renewcommand{\thefigure}{S\@arabic\c@figure}

\section{Quantum dot wafer}
The QD wafer was grown via a molecular beam epitaxy. As schematically shown in Fig.~S1(a), 300~nm GaAs buffer layer was firstly grown on a 350-$\mu$m-thick GaAs substrate followed by a 1500 nm AlGaAs sacrificial layer. 500 nm GaAs with three layers of high-density InAs QDs was finally grown on top. An atomic force microscope (AFM) image of uncapped InAs/GaAs QDs indicates a QD density of $\sim$10$^{10}$/cm$^2$ is presented in Fig.~S1(b).
\begin{figure*}[htb!]
	\begin{center}
		\includegraphics[width=1\linewidth]{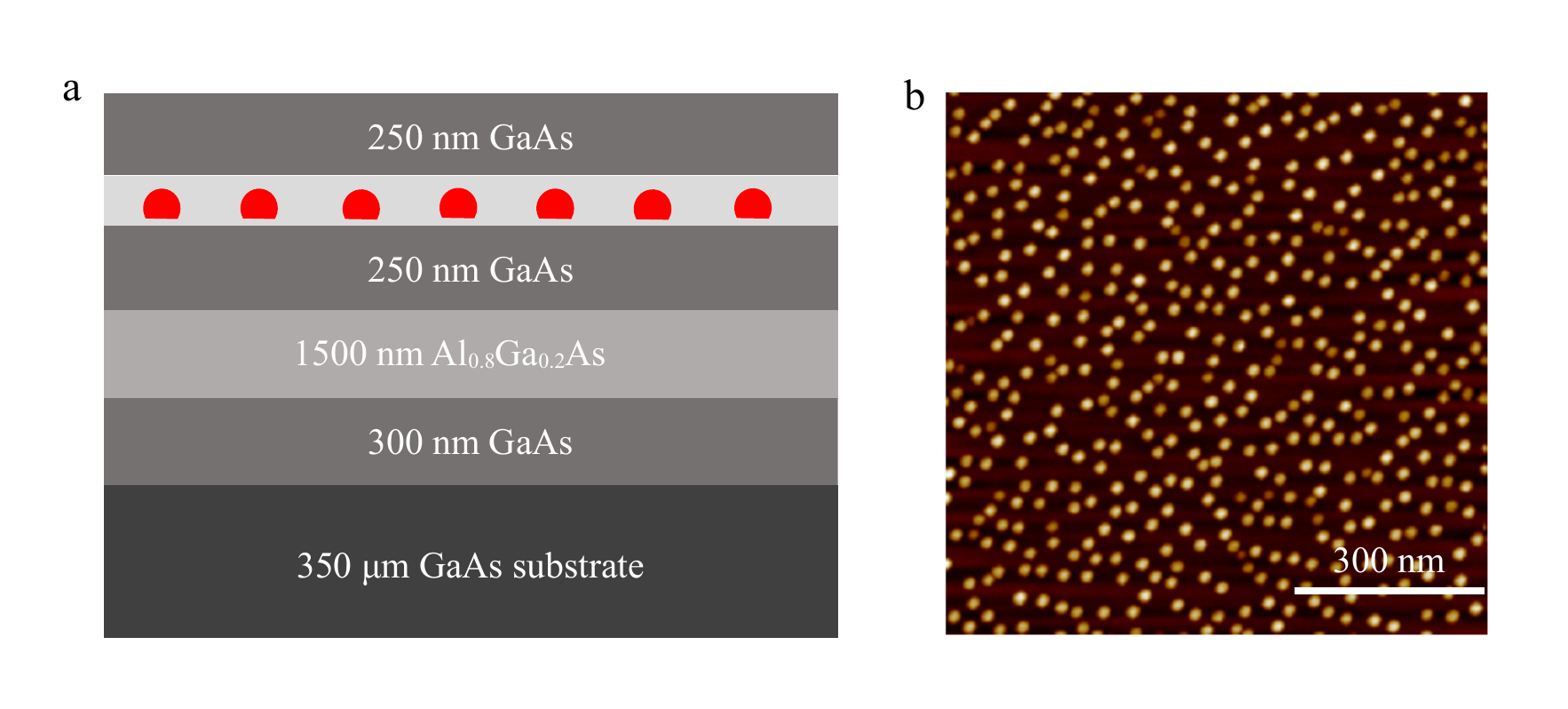}
		\caption{\textbf{QD wafer.} a, Schematic epitaxial structure of wafer for mini-BIC laser. b, AFM image of the uncapped QDs.}
		\label{fig:FigS1}
	\end{center}
\end{figure*}

\newpage
\section{Device fabrication}
The full fabrication flow of the mini-BIC laser is shown in Fig.S2. First, a layer of Si$_3$N$_4$ with a thickness of 200 nm was deposited on the wafer byiInductively coupled plasma chemical vapor deposition (ICP-CVD) as a hard mask for PhC dry etching. A 400nm ARP6200 electron beam resist was spin coated on the surface of the hard mask. Subsequently, the electron beam lithography (EBL) was used to define the PhC pattern in ARP6200. The PhC pattern was transferred from resist into the hard mask layer using reactive ion etching (RIE). Afterwards, the electron beam resist was removed by Inductively coupled plasma etching (ICP) with O2 plasma. Then a chloride-based etching was performed to obtain the air holes through the active layer. The residual Si$_3$N$_4$ hard mask was removed by RIE dry etching. Finally, the sacrificial layer was undercut by immersing the sample in 10\% hydrofluoric acid solution for 1min30s to form a suspended PhC membrane.

\begin{figure*}[htb!]
	\begin{center}
		\includegraphics[width=1.0\linewidth]{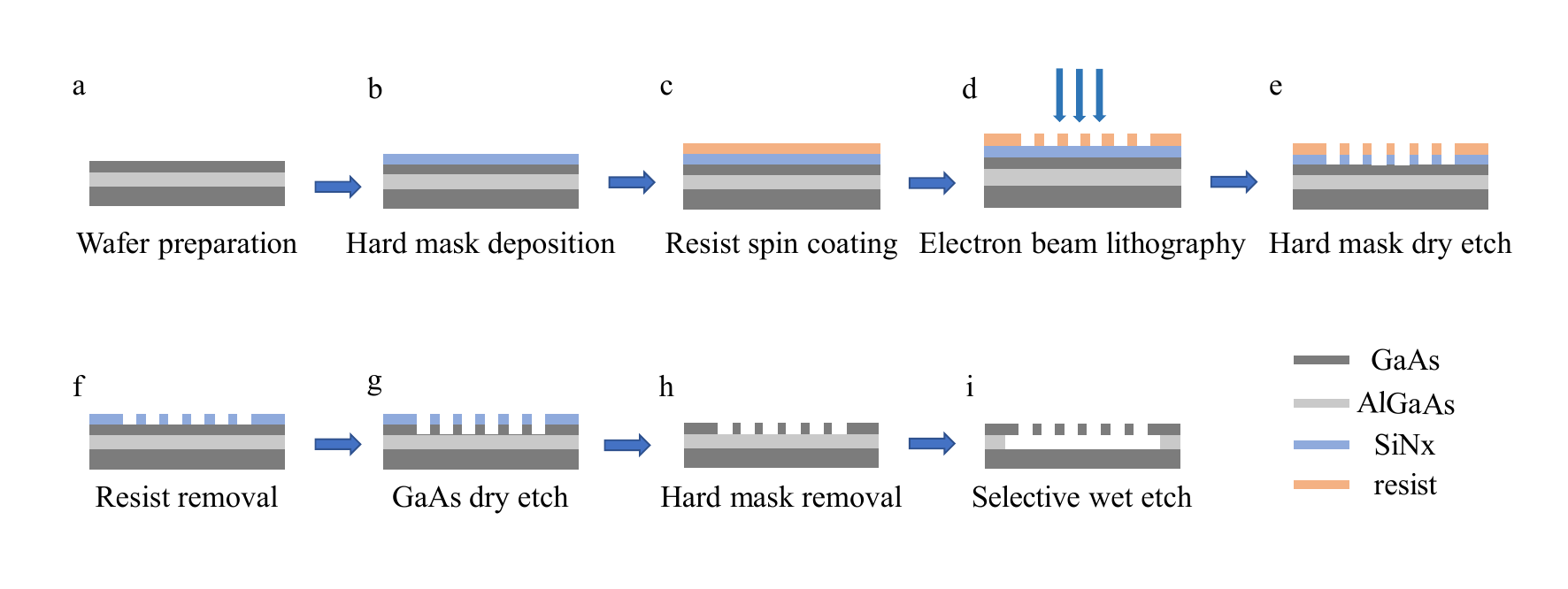}
		\caption{\textbf{Fabrication flow of the mimi-BIC laser.} a, As-grown epitaxial wafer. b, Deposition of Si3N4 hard mask. c, Spin-coating of E-beam resist as a soft mask. d, Patter definition with EBL. e, Transfering PhC pattern to hard mask layer by RIE. f, Removal of the E-beam resist. g, TransferING pattern to active layer by ICP. h, Removal of the hard mask by RIE. i, Creation of the suspended memebrane with a selective wet etching.}
		\label{fig:FigS2}
	\end{center}
\end{figure*}

\newpage
\section{Optical setup}
The cuostomied confocal $\mu$PL setup for optical characterizations is shown in Fig.~S3. The device is excited optically by a 780 nm CW laser diode or a pulse laser (10 ps, 86MHz period) through a microscope objective with a numerical aperture (NA) of 0.65. The spot size of the pump laser is about 8.11 $\mu$m, as shown in Fig.~S4. The signal emitted from the sample is collected by
the same objective, and then analyzed by a spectrometer (iHR550) with a resolution of 0.025 nm or guided to time correlated
single-photon-counting (TCSPC) measurements for the second-order correlation or lifetime measurements via a flip mirror. For
TCSPC measurements, the signal first passes through a home made grating filter with a spectral resolution of 1 nm to isolate
the laser mode. Then the filtered signal went into either one superconducting nanowire single-photon detectors (SNSPD) for
lifetime measurements or into a 50:50 fiber beam splitter (BS) and two SNSPDs for intensity autocorrelation measurements.
\begin{figure*}[htb!]
	\begin{center}
		\includegraphics[width=0.8\linewidth]{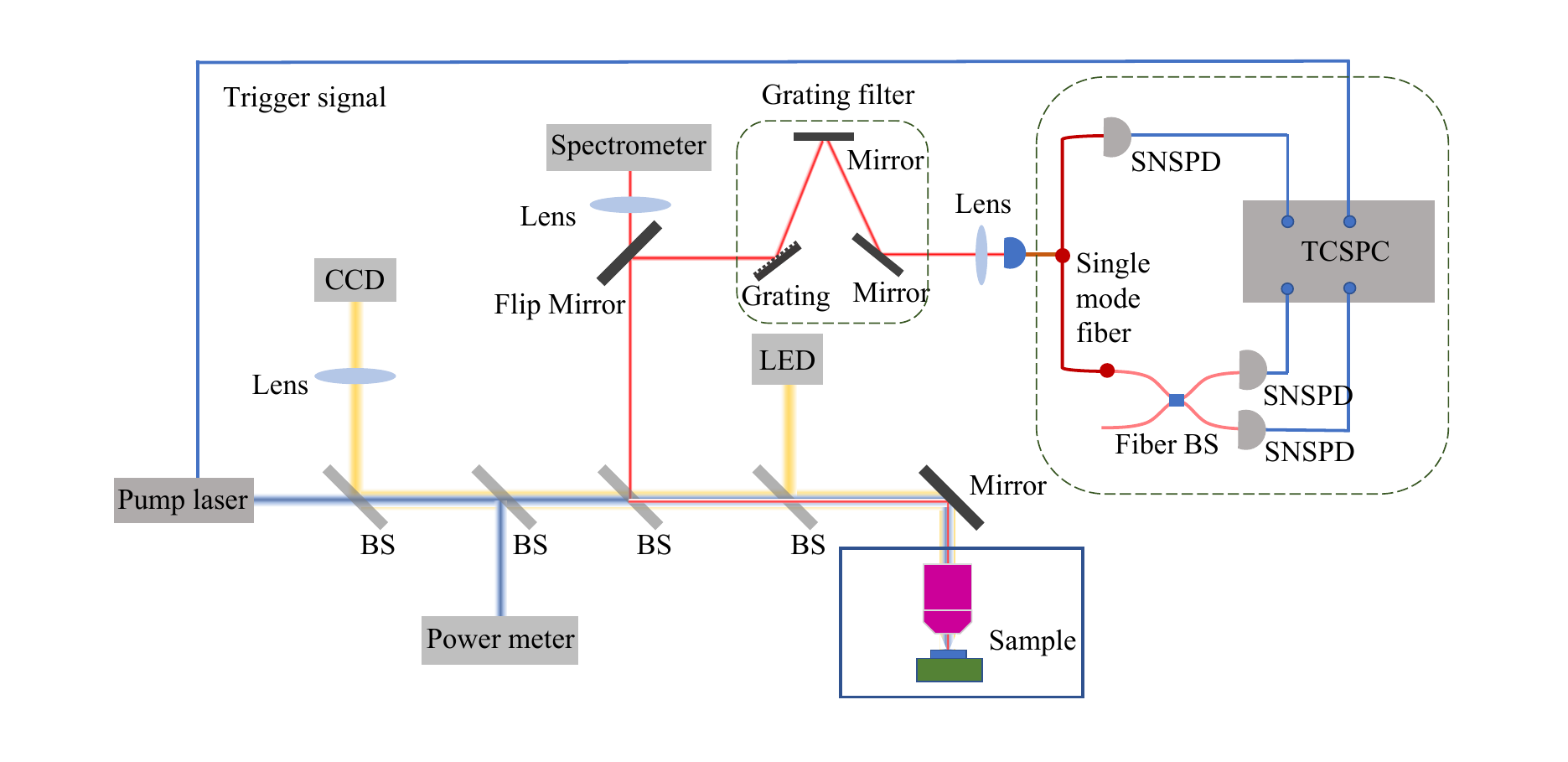}
		\caption{\textbf{Experimental setup for Optical characterization. } }
		\label{fig:FigS3}
	\end{center}
\end{figure*}

\begin{figure*}[htb!]
	\begin{center}
		\includegraphics[width=0.8\linewidth]{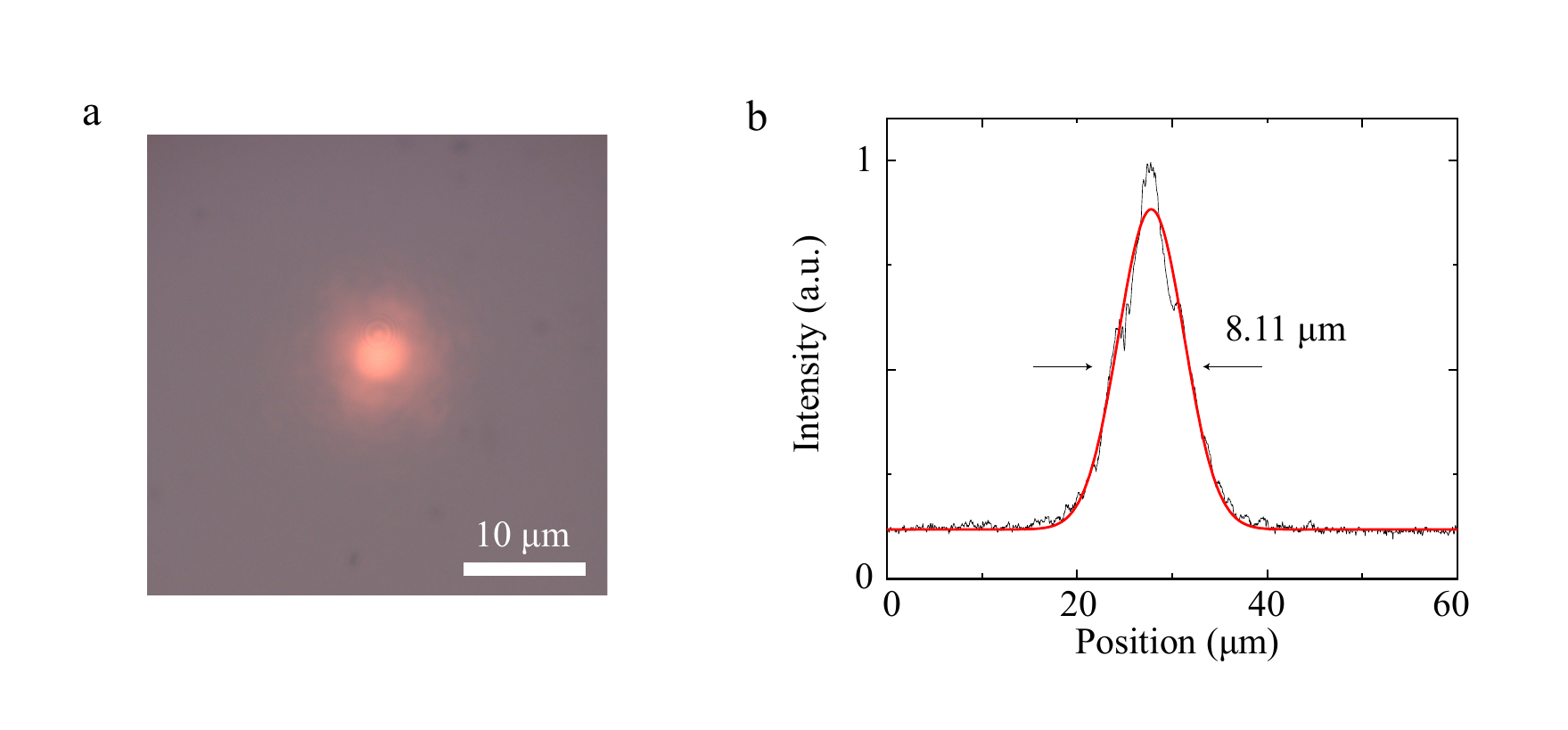}
		\caption{\textbf{The estimate of the pump spot sizes.} }
		\label{fig:FigS4}
	\end{center}
\end{figure*}

\newpage
\section{Pulsed excitation}
Fig.S5 presents the comparison of the lasing and non-lasing devices under pulsed excitation. The lasing device exhibited single-mode operation with typical lasing signatures of a nonlinear increased IO cureve and a line-width reduction behavior. Similarly, the nano-LED also showed a conventional "laser-like" behaivor in both the IO curve and line-width. 

\begin{figure*}[htb!]
	\begin{center}
		\includegraphics[width=1.0\linewidth]{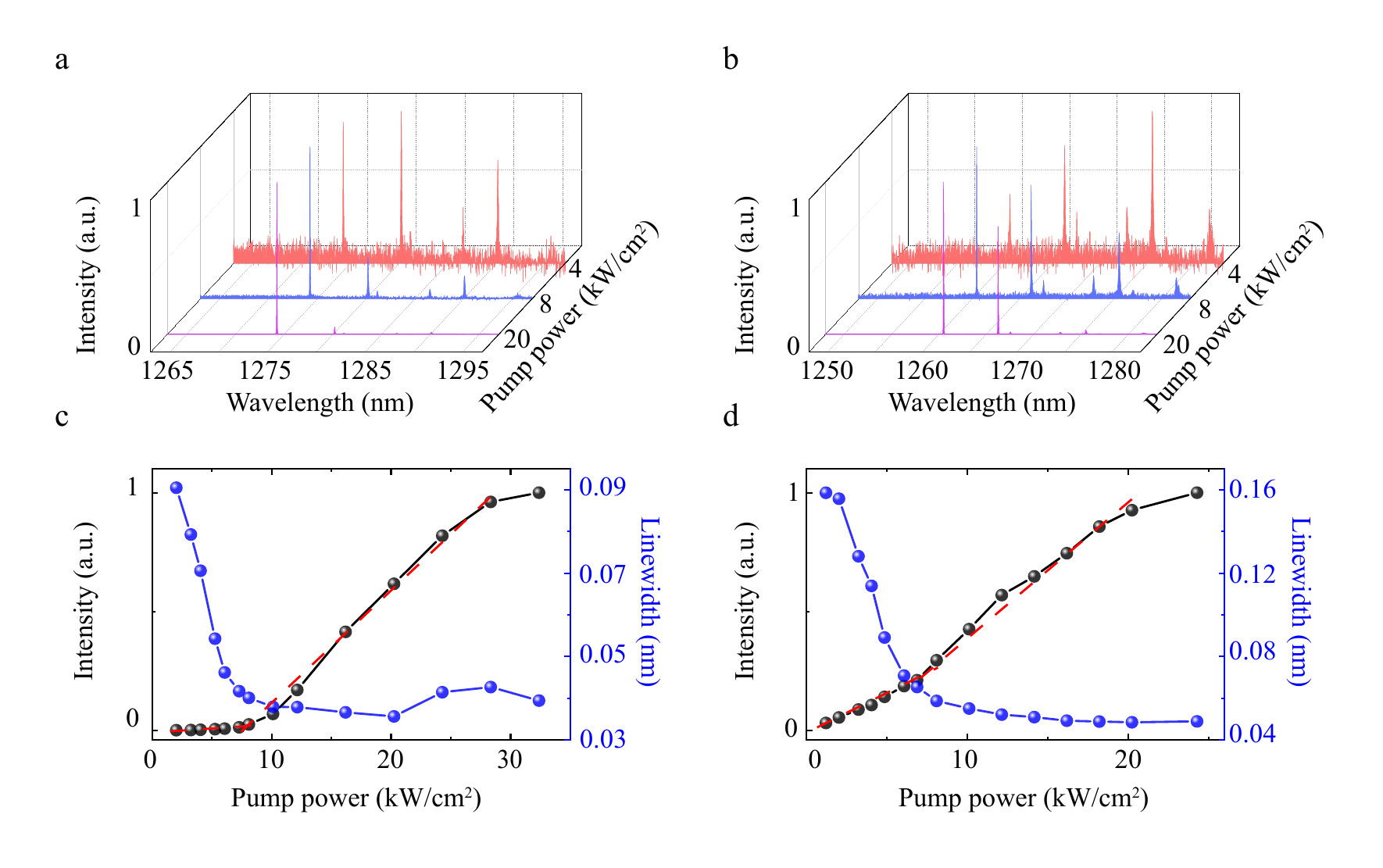}
		\caption{\textbf{Characterizations of lasing and non-lasing devices under pulsed excitation.} a,b, Evolution of the normalized emission spectra of the lasing device (a) and non-lasing device (b) with the increased excitation power. c,d, Integrated output intensity and line-width of the cavity mode M11 as a function of the excitation power, both showing “threshold-like” behaviors.  }
		\label{fig:FigS5}
	\end{center}
\end{figure*}

\newpage
\section{Comparion of mini-BIC laser with other BIC lasers}

\begin{table*}[htb!]
	\centering
	\caption{\textbf{Comparison of the mini-BIC laser with other BIC lasers.} }
%	\resizebox{\textwidth}{30mm}{
		%\setlength{tabcolsep}{35mm}{
		\begin{tabular}{llllll}
			\hline
			& Laser type & \begin{tabular}[c]{@{}l@{}}Pump \\ method\end{tabular} & \begin{tabular}[c]{@{}l@{}}Threshold \\ peak power (mW)\end{tabular} & \begin{tabular}[c]{@{}l@{}}Threshold power \\ density (kW/cm$^2$)\end{tabular} & Q factor   \\ \hline
			Nature 541, 196 (2017)     & BIC        & Pulse                                                  & 15.6                                                              & $\sim$4                                                                     & $\sim$4701 \\ 
			arXiv:1707.00181           & BIC        & Pulse                                                  & 73                                                                   & -                                                                           & -          \\ 
			Nature Nanotech. 13 (2018) & BIC        & Pulse                                                  & $8.80\times10^5$                                           & $7.0\times10^4$                                                   & 2750       \\ 
			npj 2D Materials and Applications.3 (2019) & BIC        & CW                      & -                                           & 0.144                                                   & 2500       \\ 
			Science 367 (2020)         & BIC        & Pulse                                                  & $5.28\times10^5$                                           & $4.2\times10^4$                                                   & -          \\ 
			Nano Lett. 20 (2020)       & BIC        & Pulse                                                  & $5.09\times10^8$                                           & $1.8\times10^5$                                                   & 2590       \\ 
			arXiv:2012.15642           & BIC        & Pulse                                                  & $\sim2.16\times10^6$                                     & $\sim2.75\times10^8$                                            & $\sim$2883 \\ 
			Nature Commun. 12 (2021)   & Super-BIC  & Pulse                                                  & 0.34                                                                 & 1.47                                                                        & $\sim$7300 \\ 
			\textcolor{red}{This work}                  & \textcolor{red}{Mini-BIC}   & \textcolor{red}{CW}                                                     & \textcolor{red}{0.041}                                                                & \textcolor{red}{0.08}                                                                      & \textcolor{red}{$\sim$32500}      \\ \hline
	\end{tabular}
\end{table*}

\end{document}